\documentclass[reprint,
 amsmath,amssymb,
 aps,
]{revtex4-2}
\usepackage[USenglish]{babel} 
\usepackage{amsfonts}
\usepackage{graphicx}
\usepackage{caption}
\usepackage{subcaption}
\usepackage{xcolor}
\usepackage{physics}
\usepackage{siunitx}
\graphicspath{{Images/}}
\usepackage{float}
\newcommand{\Dunit}{\si{\micro\metre\squared\per\second}}
\newcommand{\Drunit}{\si{\radian\squared\per\second}}

\newcommand{\Iunit}{\si{\micro\watt\per\micro\metre\squared}}

\definecolor{myBlue}{rgb}{0.1,0.1,0.7}

\begin{document}
\title{Run-and-tumble motion of ellipsoidal microswimmers}

\author{Gordei Anchutkin}
    \affiliation{Molecular Nanophotonics Group, Peter Debye Institute for Soft Matter Physics, Leipzig University, 04103 Leipzig, Germany}
\author{Viktor Holubec}
   \affiliation{Department of Macromolecular Physics, Faculty of Mathematics and Physics,Charles University, 18000 Prague, Czech Republic}
\author{Frank Cichos}
    \email{cichos@physik.uni-leipzig.de}
    \affiliation{Molecular Nanophotonics Group, Peter Debye Institute for Soft Matter Physics, Leipzig University, 04103 Leipzig, Germany}

\begin{abstract}
   A hallmark of bacteria is their so-called "run-and-tumble" motion, consisting of a sequence of linear directed "runs" and random rotations that constantly alternate due to biochemical feedback. It plays a crucial role in the ability of bacteria to move through chemical gradients and inspired a fundamental active particle model. Nevertheless, synthetic active particles generally do not exhibit run-and-tumble motion but rather active Brownian motion. We show in experiments that ellipsoidal thermophoretic Janus particles, propelling along their short axis, can yield run-and-tumble-like motion even without feedback. Their hydrodynamic wall interactions under strong confinement give rise to an effective double-well potential for the declination of the short axis. The geometry-induced timescale separation of the in-plane rotational dynamics and noise-induced transitions in the potential then yields run-and-tumble-like motion. 
\end{abstract}

\maketitle
Surfaces are of considerable importance for the dynamics of living active particles like bacteria. Besides the distinct run-and-tumble dynamics of, for example, {\it E.~coli}, controlled by internal bio-chemical circuitry, bacteria start to swim in circles at boundaries \cite{lauga2006swimming}, exhibit distinct spatial orientations \cite{sipos2015hydrodynamic}, or swirl in lateral confinement even with enhanced speeds \cite{wioland2013confinement}, all based on hydrodynamic interactions. Such boundary interactions may, therefore, strongly contribute to the accumulation of bacteria near surfaces, resulting in the creation of biofilms \cite{kantsler2013ciliary,wu2015amoeboid,berke2008hydrodynamic,spagnolie2012hydrodynamics,michelin2014phoretic}.
Over the last decade, synthetic active particles have provided further insights into these effects, owing to their self-propulsion mechanism, which is now often well understood and controlled. Studies on such microswimmers have unveiled complex behaviors near planar walls and obstacles, including sliding and hovering governed by pure hydrodynamics or even involving underlying phoretic mechanisms of self-propulsion \cite{takagi2014hydrodynamic,uspal2015self,das2020floor,liebchen2019interactions,simmchen2016topographical}. Despite the hydrodynamic interactions, the dynamics is largely described by a persistent random walk, where a single rotational diffusion time randomized the propulsion direction of the often spherical objects. Non-spherical particles like rods \cite{bar_self-propelled_2019,grosmann_particle-field_2020} or ellipsoids \cite{han_brownian_2006,zheng_self-diffusion_2010,shemi_self-propulsion_2018,nishigami_influence_2018} also yield a persistent random walk in their quasi-2D motion in common experimental settings. Nevertheless, irregular shapes may also provide more complex dynamics~\cite{sharan_fundamental_2021,kummel_circular_2013,chakrabarty_brownian_2013}, e.g., due to 2d-chirality, yielding circular trajectories similar to the rotational dynamics observed under circular confinements~\cite{zhang_active_2017,kummel_circular_2013}.

However, most experimental and theoretical studies on non-spherical swimmers so far consider active agents with a single orientation relaxation time, determining the transition between active ballistic and effective diffusive motion regimes \cite{bechinger_active_2016,vutukuri_light-switchable_2020, soker_how_2021}. Multiple orientational relaxation timescales that govern the transition between directed and enhanced diffusive motion may, however, yield new effects, which could be of strong relevance for discovering new collective behaviors \cite{paoluzzi_motility-induced_2022}.

Here, we experimentally study the dynamics of thermophoretically propelled Janus particle ellipsoids (JPE) strongly confined between two substrate walls. We quantitatively evaluate gravitational, optical, and hydrodynamic forces in numerical studies, comparing the results with our experimental data. We find that the confinement yields a double well potential for the polar angle of the propulsion vector due to strong hydrodynamic interactions. The resulting parallel alignment of the propulsion vector is in contrast to the wall interaction expected for puller-type flow fields of the JPE. The ellipsoidal geometry induces a timescale separation of the noise-induced transition between the double well minima and the in-plane rotational Brownian motion of the ellipsoid, which yields dynamics that is strongly reminiscent of the run-and-tumble motion of bacteria.

\begin{figure*}[!htb]
\centering    
\includegraphics[width=0.7\textwidth]{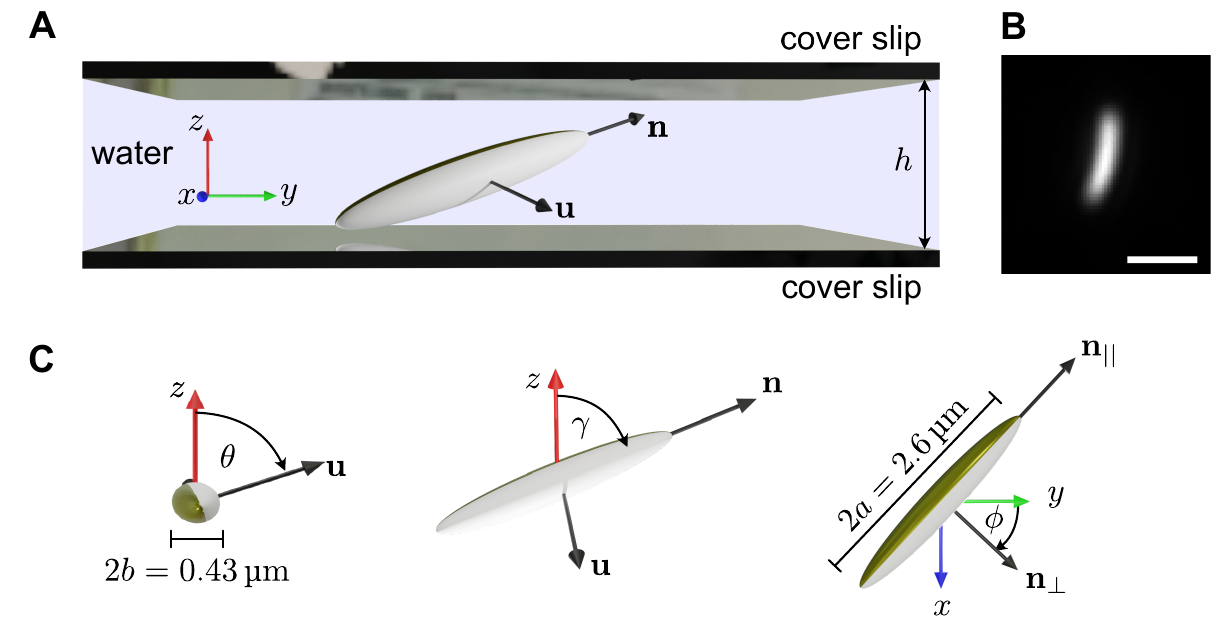}
\caption{
\textbf{A} Sketch of the sample confining a gold capped Janus-ellipsoid ($2a=2.6\, \si{\micro\metre}$ long and $2b=0.43\, \si{\micro\metre}$ short axes) in a $h=1\, \si{\micro\metre}$ or $h=2\, \si{\micro\metre}$ thin water film between two glass cover slips. \textbf{B} Example experimental darkfield microscopy image, which is used to extract the 3D orientation of the ellipsoid based on its intensity and asymmetry. \textbf{C} Definition of the angles $\theta,\gamma,$ and $\phi$ in the laboratory frame. The unit vectors $\textbf{n}_{\bot}$ and $\mathbf{n}_{\parallel}$ denote projections of the propulsion direction $\mathbf{u}$ and the vector $\mathbf{n}$ to the sample plane.
}
\label{fig:figure1}
\end{figure*}

\subsection{Experimental and data analysis}

The JPE are prepared from polystyrene ellipsoids (long axis: $2a=2.6\,\si{\micro\metre}$, short axis $2b=0.43\,\si{\micro\metre}$, aspect ratio $6$) by evaporating a $50\, \si{\nano\metre}$ gold film on half of each ellipsoid, as it is usual for Janus particles \cite{anthony_single-particle_2006,bechinger_active_2016,shemi_self-propulsion_2018}. The resulting JPE are suspended in a $1\, \si{\micro\metre}$ or $2\, \si{\micro\metre}$ thin water film between two glass cover slips (see Fig.~\ref{fig:figure1}\textbf{A}). The self-propulsion of the Janus particles is induced by a $532\, \si{\nano\metre}$ wavelength laser illuminating an area of about $60\times 60\, \si{\micro\metre^2}$ homogeneously. The illumination causes an asymmetric temperature gradient on the particle surface, inducing boundary flows that yield the particle motion in the direction $\textbf{u}$ opposite the gold cap \cite{kroy_hot_2016,bregulla_stochastic_2014,jiang_active_2010}.

The motion of the JPE is observed in a darkfield microscopy setup using an LED and an oil-immersion darkfield condenser (Olympus). The darkfield illumination causes a stronger reflection from the gold cap than from the polystyrene. The scattered light from the JPE is collected with a 100$\times$/NA\,0.6 objective lens (Olympus RMS100X-PFOD) and recorded by a camera (Hamamatsu ORCA Flash) with an exposure time of $\tau=33  \, \si{\milli\second}$, given by the LED, and an inverse framerate of $f^{-1}=33\, \si{\milli\second}$. Figure \ref{fig:figure1}\textbf{B} shows an example darkfield image used to extract the particle position as well as its 3-dimensional orientation in Fig.~\ref{fig:figure1}\textbf{C}, i.e., the angles $\theta$, $\phi$ and $\gamma$ and hence the propulsion direction $\textbf{u}$, its projection to the sample plane $\textbf{n}_{\bot}$, and the perpendicular in-plane direction $\textbf{n}_{\parallel}$.
The position change over time yields the projected displacements $\Delta_{\bot}$ and $\Delta_{\parallel}$ along $\textbf{n}_{\bot}$ and $\textbf{n}_{\parallel}$. The corresponding mean $\langle \Delta_{\bot,\parallel} \rangle$ and the variances $\langle \Delta_{\bot,\parallel}^2 \rangle$ measure the in-plane speeds $v_{\bot,\parallel}=\langle \Delta_{\bot,\parallel} \rangle \tau^{-1} $ and the diffusion coefficients $D_{\bot,\parallel}=\langle \Delta_{\bot,\parallel}^2 \rangle \left (2T-2\tau/3 \right)^{-1}$, which we corrected for the low-pass filtering and discrete sampling while imaging.


\subsection{Translational dynamics}

\begin{figure*}[!htb]
    \includegraphics[width=0.98\textwidth]{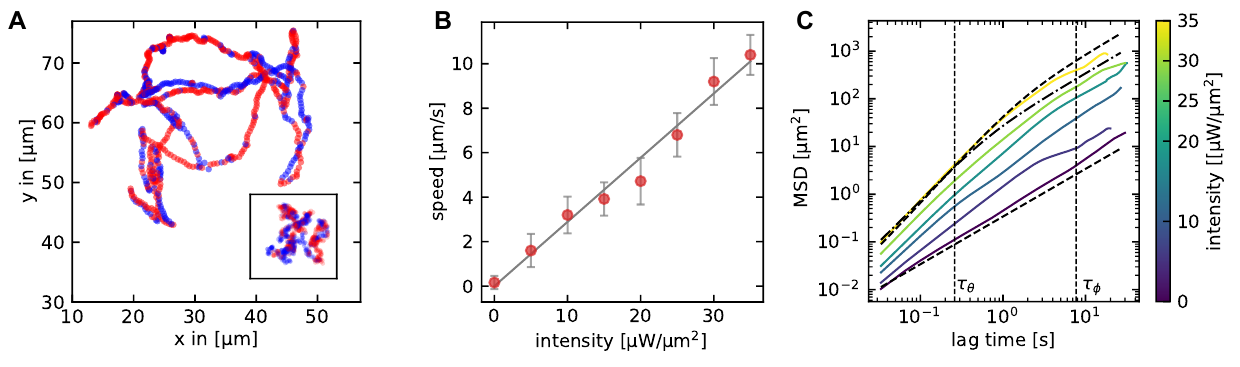}
    \centering    
    \caption{
    \textbf{A}  The recorded trajectory of the active Janus ellipsoid for an incident heating laser intensity of $I=35\, \Iunit$. The inset provides the trajectory of the passive ellipsoid ($I=0\, \Iunit$) for comparison. The color code represents the sign of the out-of-plane angle $\theta$ as defined in Fig.~\ref{fig:figure3}\textbf{A}.
    \textbf{B} Measured propulsion speed of the Janus particle ellipsoid as a function of the incident heating laser intensity. The solid line represents a linear fit to the experimental data (symbols).
   \textbf{C} Mean squared displacements (MSD) of the Janus ellipsoid for $I=0\ldots 35\, \Iunit$ reveals a "ballistic" ($\propto t^2$) and an effective diffusive ($\propto t$) regimes. The vertical lines indicate the rotational diffusion times around the long and short axis of the ellipsoid. The black lines depict analytical predictions for the passive ellipsoid (lower dashed line), freely-rotating active ellipsoid (dot-dashed line), and active ellipsoid with out-of-plane angle diffusing in a double well potential (upper dashed line), as detailed in the text.
    }\label{fig:figure2}
\end{figure*}

Figure \ref{fig:figure2}\textbf{A} depicts a characteristic trajectory of a passive ($I=0\, \Iunit$, Supplementary Video 1) and a self-propelled JPE ($I=35\, \Iunit$, Supplementary Video 2). The trajectory points are colored by the out-of-plane angle $\theta$, indicating if the asymmetrically shaped image (cf.~Fig. \ref{fig:figure1}\textbf{B}) is pointing left or right for a given orientation of the long axis. The passive dynamics (inset) exhibit rapid switches between the two orientations and can be characterized by two unique values of the diffusion tensor, which we determined to be $D_{\bot}=0.07\,\Dunit$ and $D_{\parallel}=0.16\,\Dunit$. These values can be compared to the predicted values $D_{\bot,\parallel} = k_\text{B} T (6\pi\eta b C_{\bot,\parallel})^{-1}$ for prolate ellipsoids~\cite{han2009quasi,perrin_mouvement_1936}, where $C_{\bot,\parallel}$ represents the geometric correction coefficients. In addition, the prediction needs to be corrected for the hydrodynamic wall interaction (see Supplementary Information~\cite{supplementary}), which alters the drag coefficient under strong confinement. Following \cite{li_diffusion_2004}, we obtained $D_{\bot}=0.09\, \Dunit$ and $D_{\parallel}=0.14\, \Dunit$, which agree well with our experimental results.

The active dynamics of the JPE in Fig.~\ref{fig:figure2}\textbf{A} at an incident laser intensity of $I=35\, \Iunit$ is characterized by long 'runs' with stable orientation. While the diffusion coefficient $D_{\parallel} = 0.14\, \Dunit$ along the long axis is in the active state almost unchanged, the value $D_{\bot} = 0.13\, \Dunit$ along the short axis is almost twice as high compared to the passive case. Since the active propulsion superimposes the diffusive dynamics even on the timescale of an exposure time $\tau$, we assume that $D_{\bot}$ is enhanced by activity (see Supplementary information~\cite{supplementary}). 

The self-propulsion speed $v_{\parallel}=\Delta_{\parallel} \tau^{-1}$ along the long axis vanishes as expected. The self-propulsion speed $v_{\bot}$ along the short axis increases linearly with the laser intensity as depicted in Fig.~\ref{fig:figure2}\textbf{B}. Since $\Delta_{\bot}$ is a projection of the displacements of the JPE in 3D on the in-plane direction $\textbf{n}_{\bot}$, the measured propulsion speed  $v_{\parallel}$ is smaller or equal to the actual propulsion speed of the ellipsoid.

In contrast to the well-studied dynamics of an active rod propelling along its long axis governed by only a single rotational diffusion coefficient \cite{nishigami_influence_2018}, the propulsion $\textbf{u}$ of the JPE studied here is randomized by the rotation around the short and the long axis. This is unique as compared to many other systems studied in quasi-two dimensions \cite{simmchen2016topographical}. 
The corresponding two timescales can be observed in the mean squared displacement (MSD) of the active JPE depicted in Fig.~\ref{fig:figure2}\textbf{C}. While the MSD measured for the passive JPE and active JPE in a thick sample ($h=2 \,\si{\micro\metre}$, data shown in the Supplementary Information~\cite{supplementary}) can be described by a model with a free rotational diffusion (bottom dashed and top dash-dotted lines in Fig.~\ref{fig:figure2}\textbf{C}), the MSD of the active JPE only agrees with the prediction from a dichotomic model for the JPE out-of-plane orientation $\theta$ (top dashed line), which we justify in the remainder. The derivations of the analytical expressions for the MSD are given in the Supplementary Information~\cite{supplementary}.

\subsection{Rotational dynamics}

The ellipsoid performs rotational Brownian motion around all three principle axes. The rotation of the vector $\textbf{n}_{\bot}$ in the sample plane is connected to the angle $\phi$  (cf.~Fig.~\ref{fig:figure1}\textbf{C}) and can be directly accessed in the experiment from the 2D orientation of the ellipsoid. The corresponding diffusion coefficient $D_\phi$ is determined from the Gaussian fit to the probability densities for angular displacements $\Delta \phi$ (see SI~\cite{supplementary} Fig.~S3\textbf{A}). For the passive ellipsoid $D_\phi= 0.12\, \Drunit$, which slightly increases to $D_\phi = 0.15\, \Drunit$ at an intensity of $I = 35\, \Iunit$. 
The experimental result can be compared to the theoretical prediction for a prolate ellipsoid
\begin{equation}
    D_{\phi,\theta} = \frac{k_\text{B} T }{6 \eta V C_{\phi,\theta}}
    \label{eq:Dr}    
\end{equation}
with the ellipsoid volume $V$, the viscosity $\eta$, and a geometric factor for the corresponding axis of rotation $C_{\phi,\theta}$, which reflects the deviation from a sphere~\cite{perrin_mouvement_1936,koenig_brownian_1975}.
To correct this prediction for the increased apparent friction due to the confinement is in 3D rather complicated since the JPE can diffuse in the $z$ direction and rotate along the angle $\gamma$ (cf.~Fig.~\ref{fig:figure1}\textbf{C}). Following the calculations for a cylindrical rod in \cite{li_diffusion_2004}, we obtained the analytical estimate $D_\phi = 0.14\, \Drunit$, which is in fair agreement with the experimentally determined result.

\begin{figure}[!htb]
    \includegraphics[width=0.9\columnwidth]{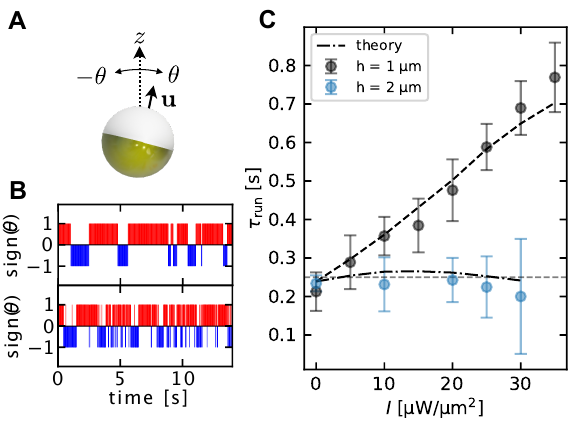}
    \centering    
    \caption{
    \textbf{A} Definition of negative and positive angles $\theta$ when keeping the orientation of the long axis $\mathbf{n}$ of the ellipsoid fixed. 
    \textbf{B} The sign angle trace $\theta(t)$ as a function of time as extracted from the experiment for $h=1\,\si{\micro\metre}$ (top) and $h=2\,\si{\micro\metre}$ (bottom). 
    \textbf{C} The measured (circles) and theoretically predicted (black lines, details in Supplementary Information~\cite{supplementary}) switching time $\tau_{\rm run}$ for $h=1\,\si{\micro\metre}$ (black circles, dashed line) and $h=2\,\si{\micro\metre}$ (blue circle, dot-dashed line) as a function of the heating laser intensity. The gray horizontal line indicates the rotational diffusion time around the long axis $\tau_{\theta}$ for the passive ellipsoid.
    }
    \label{fig:figure3}
\end{figure}

The rotational motion of the JPE around the long axis $\textbf{n}_{\parallel}$ can be determined from either the shape asymmetry of its image or the ellipsoid's scattering intensity difference $\Delta I_{\rm S}$ under darkfield illumination \cite{anthony_single-particle_2006,behrend2004metal,qian2013harnessing,bregulla_stochastic_2014}. 
The intensity difference is minimal when the gold cap faces upwards ($\theta=180\, \si{\degree}$) and maximal for $\theta=0\,\si{\degree}$ (cf.~Fig.~\ref{fig:figure1}\textbf{C}). At intermediate orientations, we map $\Delta I_{\rm S}$ to the angle $\theta$ approximately as $\Delta I_{\rm S} = \Delta I_0(1+\cos\theta)/2$ ~\cite{anthony_single-particle_2006,behrend2004metal}, as detailed in the Supplementary Information~\cite{supplementary}. The slow rotation around the short axis of the ellipsoid allows us further to discern the orientation of the curved-shaped image of the JPE (cf. Fig.~\ref{fig:figure1}\textbf{B}) for a given orientation of the long axis. We define the corresponding angle $\theta$ as positive or negative if the vector $\textbf{n}_{\bot}$ points either to the left or right for a specific orientation of $\textbf{n}_{\parallel}$. The time series $\theta(t)$ or $\operatorname{sign}(\theta(t))$ as displayed in Fig.~\ref{fig:figure3}\textbf{B} can thus be used to inspect the rotational dynamics around the long axis, e.g., by measuring the average switching time $\tau_{\textrm{run}}$ between positive and negative values of $\theta$. 

\begin{figure*}[!ht]
    \includegraphics[width=0.9\textwidth]{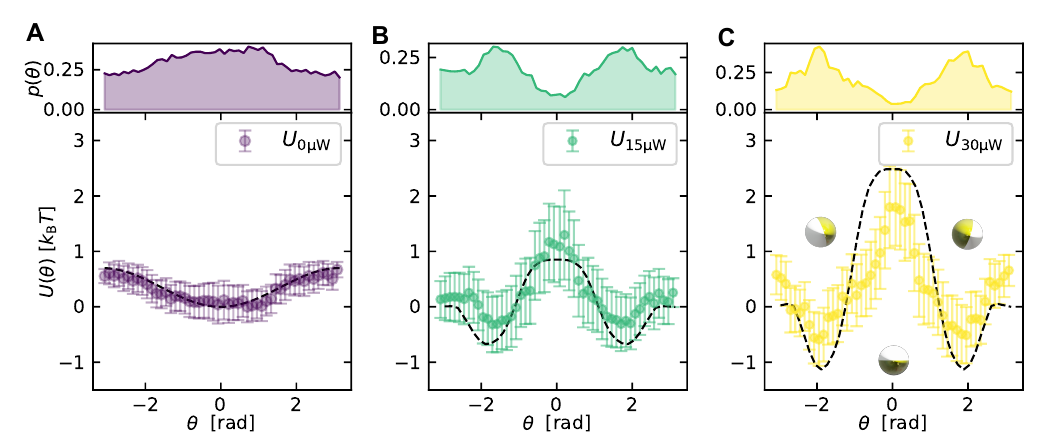}
    \centering    
    \caption{Probability density $p(\theta)$ and potential of mean force $U(\theta)$ for the angle $\theta$ between the propulsion direction and the z-axis at the laser intensities of \textbf{A} $I = 0\, \Iunit$, \textbf{B} $I = 15\, \Iunit$, and \textbf{C} $I = 30\, \Iunit$. The angular distributions (upper part) are calculated from the particle orientation traces $\theta(t)$ and corrected for the angular degeneracy (see SI~\cite{supplementary}). 
    The corresponding potentials of mean force (lower part) are compared to the model (dashed lines), including gravitational, optical, and hydrodynamic forces, as detailed in the text. 
    }
    \label{fig:figure4}
\end{figure*}

Figure \ref{fig:figure3}\textbf{C} shows that for a passive ellipsoid the switching times $\tau_{\textrm{run}}$ coincide for both film thicknesses and are close to the rotational diffusion time $\tau_{\bot}=D_\theta^{-1}=0.26\,\si{\second}$ around the ellipsoid's long axis (horizontal dashed line in Fig.~\ref{fig:figure3}\textbf{C}). In the thicker sample ($h=2\,\si{\micro\metre}$), the switching time is nearly independent of the laser intensity. Under the strong confinement in the thin film ($h=1\,\si{\micro\metre}$), it increases with increasing power. 

To understand the intensity dependence of the switching dynamics of $\theta(t)$ in Fig.~\ref{fig:figure3}, we depict in Figs.~\ref{fig:figure4}\textbf{A--C} the probability density for $\theta$ for three different laser intensities. In the passive case, the distribution has a broad maximum around $\theta=0\, \si{\radian}$, corresponding to the gold cap pointing downwards. At intermediate and the highest intensities of $I = 15\, \Iunit$ and $I = 30\, \Iunit$, the histogram is bimodal with a minimum at $\theta=0\, \si{\radian}$ and two maxima located at $\theta=\pm 2\, \si{\radian}$ or $115\,\si{\degree}$ for the highest power. The corresponding orientations of the gold cap are sketched in Fig.~\ref{fig:figure4}\textbf{C}. The bottom panels in Fig.~\ref{fig:figure4} show effective potentials of mean force for $\theta$ calculated as $U(\theta) = -\log p(\theta)$. 

\subsection{Gravity, Optical and Hydrodynamic Forces}

\begin{figure}[!htb]
    \includegraphics[width=0.9\columnwidth]{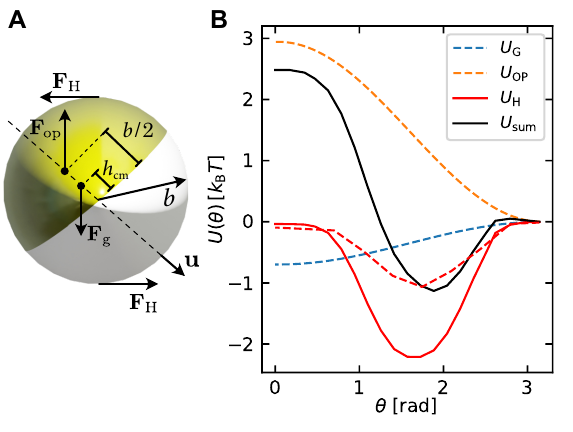}
    \centering    
    \caption{\textbf{A} Forces acting on the JPE due to radiation pressure ($\textbf{F}_{\rm op}$), gravity ($\textbf{F}_{\rm g}$) and hydrodynamic forces ($\textbf{F}_{\rm H}$). $b$ denotes the half-length of the short axis. The center of mass (COM) is displaced by $h_{\rm cm}$ from the geometrical center. The optical forces act on a center displaced by $b/2$ from the center of geometry. Hydrodynamic forces act at the boundary of the particle. \textbf{B} Calculated angular dependence of the potential contributions corresponding to the forces in \textbf{A}: gravity $U_{\textrm{G}}$ (blue dashed), radiation pressure $U_{\textrm{OP}}$ (green dashed), hydrodynamics $U_{\textrm{H}}$ (red solid for $h=1\, \si{\micro\metre}$, red dashed $h=2\, \si{\micro\metre}$), and total potential $U_{\textrm{sum}}$ (black solid for $h=1\, \si{\micro\metre}$). 
    }
    \label{fig:figure5}
\end{figure}

The shape of the potential in Fig.~\ref{fig:figure4}\textbf{C} results from the forces acting on the JPE, depicted in Fig.~\ref{fig:figure5}\textbf{A}. These are gravity $\textbf{F}_{\textrm{g}}$ , optical forces $\textbf{F}_{\textrm{op}}$ and hydrodynamic forces $\textbf{F}_{\textrm{H}}$. In the passive case, only gravity acts to sediment and reorients the bottom-heavy JPE.  
The gravitational torque on the particle aligns the unit vector $\mathbf{u}$ with the z-axis ($\theta=0$). The corresponding potential for $\theta$ depends on the displacement of the center of mass from the center of geometry $h_\textrm{cm}$ (see Fig.~\ref{fig:figure5}\textbf{A}) and the particle's mass $m$ (see Sec.~5 in SI~\cite{supplementary}). The potential varies as $\Delta U_{\rm g}(\theta) = -m g h_\textrm{cm}(1+\cos(\theta))=\Delta U_{\textrm{g},0}(1+\cos(\theta)) $. It is plotted using $h_\textrm{cm}=0.1 \, \si{\micro\metre}$ and $m = 1.8 \times 10^{-15} \, \si{\kilo\gram}$ in Figs.~\ref{fig:figure5}\textbf{B} and Fig.~\ref{fig:figure4}\textbf{A}, where it nicely overlaps with the potential of mean force for a passive bottom-heavy JPE determined experimentally.

The optical force $\textbf{F}_\mathrm{op} = \textbf{F}_\mathrm{sca} + \textbf{F}_\mathrm{abs}$ stems from the radiation pressure due to scattering and absorption of the incident green laser light by the gold cap. It points along the wavevector of the incident light ($\textbf{F}_\mathrm{op} \parallel \hat{\textbf{e}}_{\textrm{z}}$) and its magnitude
$F_\mathrm{op}=  F_\mathrm{sca} + F_\mathrm{abs} = R\frac{2I}{c}A + \frac{I}{c}\sigma_{\rm abs}$
is determined by the light intensity $I$, the speed of light $c$, the area $A$ of the gold cap, the reflection coefficient $R$, and the absorption cross-section of the gold cap $\sigma_{\rm abs}$~\cite{mousavi2019clustering,jones_marago_volpe_2015}, see SI for details~\cite{supplementary}. The optical force acts on the center of scattering, which is mainly defined by the optical properties of the gold cap, and aligns the gold cap upwards against the gravitational force. The potential thus has a similar shape as the gravitational potential. It has a maximum at $\theta=0\,\si{\radian}$ and decreases as $\Delta U_{\textrm{op}}(\theta)=\Delta U_{\textrm{op},0}( 1+\cos(\theta))$, with $\Delta U_{\textrm{op},0}= F_\mathrm{op} b/2$, where $b$ is the half-length of the ellipsoid's short axis. 

Since the gravitational and optical potentials just differ in magnitude and sign, they cannot reproduce the double well potentials found in the experiments. The double well shape, therefore, must result from the hydrodynamic interactions of the swimming JPE with the confining walls. Such hydrodynamic interactions for microswimmers have been studied analytically, numerically \cite{spagnolie2012hydrodynamics, michelin2014phoretic,das2020floor, Winkler_rheotaxis_2020,  liebchen2019interactions} 
and experimentally \cite{berke2008hydrodynamic,simmchen2016topographical,takagi2014hydrodynamic}. In proximity to a single wall, pushers typically align parallel to the wall and pullers perpendicular to it~\cite{spagnolie2012hydrodynamics}. In the present case of strong confinement by two walls, these insights based on the far-field description of the swimmer cannot be easily applied. Therefore, we have numerically evaluated the hydrodynamic torque on the JPE by finite element simulations using COMSOL Multiphysics v.~6.1. 

As our ellipsoid has a large aspect ratio of about 6, we carried out 2D simulations describing the dynamics of an infinite cylinder parallel to the two confining walls (see SI~\cite{supplementary} for details). We integrated the resulting hydrodynamic stress tensor along the surface of the cylinder to obtain the torque on the JPE swimming in the middle of the water film at different angles $\theta$. From this, we calculated the hydrodynamic contribution to the potential $U_{\rm H}$ in Fig.~\ref{fig:figure5}\textbf{B}. It has a minimum at $\theta=\pi/2\,\si{\radian}$, as anticipated from the symmetry of the problem. The hydrodynamic torque thus aligns the swimmer propulsion direction $\textbf{u}$ parallel with the walls. Our swimmer resembles a puller (see flow field in SI~\cite{supplementary}). Therefore, this result is in stark disagreement with the far-field prediction. 

Combining the hydrodynamic, gravitational, and optical contributions, we find a total potential with a maximum at $\theta=0\,\si{\radian}$ and a minimum at $\theta=1.9\,\si{\radian}$. The slight shift of the minimum of the hydrodynamic contribution from $\theta=\pi/2\,\si{\radian}$ and the heights of the two barriers (at $\theta = 0$ and $\theta = \pm\pi$) between the positive and negative values of the angle $\theta$ are caused by a combined effect of the gravitational and optical torques.
The resulting total potentials $U(\theta)$ depicted in Figs.~\ref{fig:figure4}\textbf{B--C} without any fitting parameter nicely agree with the experimental data. The theoretical predictions for the switching times $\tau_\text{run}$ between negative and positive angles $\theta$ calculated from the total potentials (see SI~\cite{supplementary} for details) are illustrated in Fig.~\ref{fig:figure3}\textbf{B} by dashed and dot-dashed lines. 

The interplay of the optical and hydrodynamic torques under strong confinement thus results in a bistable orientation of the angle $\theta$. As the rotation around the short axis of the ellipsoid has a very long rotational diffusion time of $\tau_{\phi}=1/D_{\phi}=7.14\, \si{\second}$, a strongly confined and illuminated ellipsoid carries out long runs with hardly any change in the in-plane orientation while trapped in one of the meta-stable $\theta$ states. These one-dimensional runs switch their direction as $\theta$ jumps to the other minimum of the double well potential with a switching time $\tau_\text{run}$ about four times larger than the rotational diffusion $D_\theta$ around the long axis. The interplay of rotational diffusion anisotropy and hydrodynamic alignment thus induces a run-and-tumble-like motion of the JPE, as shown in Supplementary Video 2. Compared to bacteria, where the tumble is actively controlled by internal biochemical feedback, the observed tumbling behavior for the ellipsoid is purely passive and induced by the effective bistability of the rotational diffusion.

In summary, we have shown that strongly confined self-propelled ellipsoidal Janus particles undergo a run-and-tumble-like motion when propelled along their short axis. Their dynamics is characterized by long straight runs where the particle's velocity only switches its sign. This is caused by a hydrodynamic alignment of the short propulsion axis parallel to the wall under the strong confinement despite the puller-type hydrodynamic far field of the swimmer. We quantitatively evaluated the gravitational, optical, and hydrodynamic forces and found that our numerical results and the experimental data are in good agreement. While spherical particles would show similar wall alignment, the slow rotational diffusion around the short axis of the ellipsoid in the present case separates the tumble time (rotation around the long axis) from the rotational diffusion time (rotation around the short axis), yielding the run-and-tumble motion. 
As the shape-related steric and hydrodynamic interactions are two important components for the collective behavior of active particles, we expect that the modified orientational dynamics and the timescale separation arising from shape asymmetry will lead to distinct motility-induced clustering. We further suggest that this interplay of shape and hydrodynamics will play an important role in the formation of non-equilibrium structures of active Janus particle ellipsoids in mixtures with other active and passive components to form new functional active systems.

\paragraph{Acknowledgement.}
GA and FC acknowledge financial support from the German Research Foundation (Deutsche Forschungsgemeinschaft, DFG) through project no 432421051.
VH acknowledges the support of Charles University through project PRIMUS/22/SCI/009. We acknowledge help with the hydrodynamic COMSOL calculations by LaiLai Zhu (NSU, Singapore) and Ján Šomvársky (MFF UK, Prague).  The authors also acknowledge the careful proofreading of the manuscript by Andrea Kramer.

\bibliography{literature}

\end{document}